# Kinetics of Bias stress and Bipolaron Formation in Poly(thiophene)

A. Salleo and R. A. Street

Palo Alto Research Center, 3333 Coyote Hill Road, Palo Alto, CA 94304

**Abstract**

Studies of the kinetics of the bias-stress effect are reported for PQT-12 polymer thin film transistors. We associate the bias-stress effect with bipolaron formation, based on the observation that the stressing rate is proportional to the square of the hole concentration. The bipolaron formation rate is measured as a function of temperature, and the thermal dissociation rate is also deduced from the data. We discuss the hole capture process and suggest that tunneling through the Coulomb barrier dominates. The temperature dependence of bipolaron formation and recovery kinetics leads to a maximum in their formation at about 220 K, and we also find other changes in the electrical properties near this temperature. A separate bias-stress effect is observed to operate at much longer time scales, and might be associated with a different population of bipolarons.

Contact: asalleo@parc.com

## 1. Introduction

Recent studies of bias-stress effects in polymer thin film transistors (TFT), found a reversible decay in the hole current due to trapping in the polymer semiconductor.[1] The rate of loss of channel charge is proportional to the square of the hole density in the accumulation layer. This result indicated that self-trapped hole pairs (bipolarons) are formed, and are responsible for at least one component of the bias-stress effect. The data compared poly(thiophene) and poly(fluorene) semiconductors and found similar mechanisms. However, different bipolaron binding energies in the two materials gave a large difference in the stress effect and the resulting TFT characteristics. Further studies showed that the stress was reversed by light absorbed in the polymer semiconductor, supporting the hole pairing interpretation.[2]

The formation of bipolarons is a general and well-known phenomenon in doped conductive polymers. Most past studies were aimed at observing the conversion of polarons into bipolarons over timescales much longer than in this work. To the best of our knowledge, the relative drift mobility of polarons and bipolarons in the same material has not been directly compared. At high doping levels (up to tens of percent unit charge per thiophene ring), the number of free spins in doped polymers is not always correlated to the conductivity leading to the conclusion that bipolarons are at least partially responsible for charge transport.[3-6] In these conditions metal-like features are observed in the absorption spectra.[7] Brédas et al.[8] have proposed that bipolaron transport dominates at high doping levels, where the otherwise localized bipolaron states form bands. In the TFT channel however only a fraction of a percent of the thiophene rings carry a charge, which is equivalent to a low doping level. Moreover, bipolarons are negative correlation energy states that must be stabilized by a large relaxation of the polymer backbone from the unpaired polaron state to the bound state. The relaxation energy depends strongly on the local structure of the polymer. In high-mobility polycrystalline regio-regular poly(thiophene) films such as those studied here, where the mobile species consists of holes partially delocalized in the $\pi$-$\pi$ stacking direction,[9] it is highly likely that bipolarons have a much smaller mobility than the unpaired holes.

The previous studies[1] measured the hole trapping kinetics at room temperature. In order to improve understanding of the mechanism of hole capture, and the reverse dissociation of the bipolarons, more information about the kinetics and its temperature dependence are needed, and this is the focus of the present paper. The formation of the hole pair or bipolaron, $(hh)_{BP}$, and the reverse dissociation, are described by the following reaction and kinetics,

$$h + h \leftrightarrow (hh)_{BP}, \qquad (1)$$

$$\mathrm{d}N_h/\mathrm{d}t = - k\, N_h(t)^2 + bN_{BP}(t) \qquad (2)$$

$$N_{h0} = N_h(t) + N_{BP}(t)/2 \qquad (3)$$

where $N_h$ and $N_{BP}$, are the hole and bipolaron concentration, and $k$ and $b$ are corresponding rate constants, both of which are expected to be temperature dependent. When the transistor is in the off-state, $N_{BP}$ =0, and eq. 1 predicts that when the device is turned on, $N_h$ and therefore the TFT current, decays from an initial value, $N_{h0}$, to a steady state which depends on the values of $k$ and $b$. The present measurements use this property to obtain information about both rate parameters.

The poly(thiophene) semiconductors are disordered, polycrystalline materials, and therefore the properties of the trapped hole pairs may depend strongly on specific location in the material, giving alternative sites or a distribution, as discussed previously.[1] Generalizations of equations 1-3 would be required to include such distributions. Furthermore, the TFT transport data give no direct information about the structure of the hole pair. Recent calculations by dos Santos suggests that the low energy state for the bipolaron occurs with the holes occupying adjacent thiophene rings in the π-π stacking direction, rather than both holes occupying the same ring.[10] A pair of holes in this structure is also referred to as a polaron pair. Our description of the hole pair as a bipolaron does not differentiate between this or any other possible pair structure.

## 2. Measurements

Measurements were made using poly[5,5'-bis(3-alkyl-2-thienyl)-2,2'-bithiophene)] (PQT-12) a poly(thiophene) synthesized by Xerox Research Centre of Canada as the semiconductor.[11,12] The material has symmetrical alkyl side chains, thus conferring regioregularity on the polymer. The fabrication of the TFT devices was described in the previous papers.[1,2] The bottom gate coplanar TFTs have gold source and drain contacts, and either a silicon thermal oxide gate dielectric, or 20 nm plasma-enhanced chemical vapor deposition (PECVD) silicon oxide on 200 nm nitride gate dielectric. The semiconductor was deposited by spin coating to thickness 30-50 nm, on a self-assembled monolayer (SAM) of octadecyl-trichlorosilane (OTS).

Pulsed and quasi-static measurements are used to characterize the transistors. Quasi-static data is obtained by sweeping the gate voltage from positive to negative voltages in 0.5 - 1V steps, in 1 - 5 s. intervals. Pulsed current measurements are captured on a digital oscilloscope, and normally the gate voltage is pulsed while the drain voltage is held constant, with measurements usually made in the linear regime. Room temperature measurements are usually performed in air or nitrogen, while temperature dependent measurements are made in vacuum.

## 3. Results

Figure 1 shows a typical TFT transfer characteristic measured at room temperature. The field effect mobility is in the range 0.05-0.1 $cm^2/V\ s$. The on/off ratio is $>10^6$, the turn-on voltage is close to 0 V and the threshold voltage is 5-10 V. The 300 K transistor data are described reasonably well by a constant mobility above the sub-threshold region. Lower temperature transfer data taken with the pulsed technique are shown in Figure 2, and exhibit an increasingly large sub-threshold region as discussed further below. The capacitative switching transient extends to longer time as the temperature is reduced and limits the measurement to about 4 ms. after the gate pulse at low temperature.

The quasi-static measurements show negligible room temperature hysteresis, when the same data are measured in quick succession, as indicated in Figure 1. However, pulsed measurements tend to give a slightly higher mobility, indicating stress effects that are rapidly reversed.

## 3.1 Long time bias stress effects and recovery

Previous studies of the stress effect focused on the current decay in the first few seconds after the gate bias is applied.[1,2] Figures 3 and 4 show the change in drain current for a TFT measured over an extended period of time at different gate voltages and temperatures. There is an initial behavior at times less than about 10 s. followed by a longer time behavior with different characteristics. The initial decay occurs rapidly and with increasing strength as the gate voltage is increased, and is the main focus of this paper. The long time behavior occurs at $10^2$ – $10^5$ s. at room temperature, also with a rate that increases with gate voltage. There is a reduction in the rate of current decay as the temperature is reduced.

The two regimes of stress effects are also distinguished by the recovery rate. When stress is applied for a short time (i.e. up to 100 s.), the recovery takes only a few seconds, as discussed in a previous publication. The evidence for the rapid recovery is that there is no observable change in the transfer characteristics when the measurement is repeated within a short time of about one minute (see Figure 1). However, the longer term stress effect recovers much more slowly. As previously observed with F8T2, we verified that the longer term stress effect is due to a shift of the threshold voltage and there is little change in the sub-threshold region of the transfer curve and no decrease in carrier mobility, as is shown by the data in Fig. 5a and b.

We measured the recovery rate of the longer term bias stress. Here the sample is stressed for 2 hours at 50 V gate bias. The recovery is allowed to proceed at zero bias and is measured by tracking the turn-on voltage $V_{on}$ of the transistor.  The recovery of the threshold voltage is shown in Fig. 6. The form of the recovery is approximated by a double exponential with characteristic time constants of approximately 1300 s and 32000 s.

Both the short-term and long-term decay are relatively independent of the choice of gate dielectric, and results for $SiO_2$ thermal oxide and PECVD silicon oxide/nitride are shown in Figure 7.  The two regimes of bias stress are therefore likely to be due to trapping at different sites in the polymer as discussed later.  In the following sections we focus on the short time stress effect only.

3.2 Bipolaron kinetics from pulsed measurements

The previous studies showed that the initial decay of the current reflects the trapping of hole pairs.[1]  The capture of holes to form the bipolaron has very low capture cross section, which was attributed to the mutual Coulomb repulsion of the holes.  The Coulomb barrier may be overcome by thermal activation or by tunneling and these mechanisms are expected to be distinguished by their temperature dependence.

As in the previous work, the rate of bipolaron formation is obtained by measurement of the initial current decrease in a pulsed measurement of the TFT, at a short enough time that the reverse process can be neglected. The measurement is performed in the linear regime of the TFT, so that the hole concentration can be assumed proportional to the current, and $dN_h/dt$ can be measured.  The decay is measured at short times to minimize the reverse process of bipolaron dissociation in the kinetics described by eq. 2, so that the rate constant $k$ is measured directly.  The hole concentration is calculated from the gate capacitance, using the relation $N_h = C_G(V_G - V_T)$, while previously it was calculated from the TFT current and from the measured mobility.  Basing the values on the gate capacitance avoids needing to know the temperature dependence of the mobility, but includes immobile holes which represent an increasing fraction of the holes when the temperature is reduced, as indicated by the TFT characteristics in Figure 2.

Figure 8 shows one example of the pulsed data and the measured value of $dN_h/dt$ as a function of $N_h^2$. The data all follow the square law form of eq. 2 and the rate constant is determined from the slope. The measurement assumes a uniform hole concentration in a channel of depth 1 nm. Given the structure of PQT-12, most of the current flows in a single thiophene plane adjacent to the interface, and so the calculated bulk density is an approximation.

Figure 9 shows the resulting temperature dependence of the apparent rate constant, for temperatures between room temperature and 140K. The room temperature rate is $3\times10^{-19}$ $cm^3$/s., compared to the previously reported value of $1.3\times10^{-18}$ $cm^3$/s. The factor 3 difference is largely due to a numerical error in the previous analysis**.** The apparent bipolaron formation rate is independent of temperature from 300 – 220 K, and at lower temperature has an activation energy of ~0.12 eV.

The hole capture rate, $k$, in eq. 2 determines the initial decay of the current while the dissociation rate, $b$, determines the steady state hole concentration. Hence, information about the bipolaron dissociation rate was obtained by analyzing the time dependence of the TFT current as a function of temperature. Figure 10a shows examples of the relative decay of the TFT current over a time period of 0.1 s., at different temperatures. There is an initial decay which tends to a steady state, and both the time dependence and the magnitude of the decay are temperature dependent. There is clearly a more pronounced current decay between 220 K and 260 K than at higher or lower temperatures. Figure 10b is a series of calculated decay curves according to the model discussed in section 4.2. The curves shown in Fig. 10b show a good qualitative similarity to the data (Fig. 10a).

From data as in Fig 10a, the ratio of the current at 0.1 sec to that extrapolated to zero time, is used as a convenient measure of the decay, and results are shown in Fig. 11. This ratio decreases as the temperature is reduced down to about 220 K and then increases again, until essentially no decay is observed in this time range at 140 K. The results imply that the current exhibits its fastest decay at 220 K, and the same result is also evident in Fig. 10. The form of Fig. 11 results from a balance between the increasing equilibrium bipolaron concentration at lower temperature and the decreasing rate of formation. The solid line is a fit to the data and the parameters of the fit give information about the bipolaron kinetics.

3.3 Pulsed and DC measurements of TFT characteristics

The effect of the bias-stress, and its variation with temperature, should be reflected in the quasi-static and pulsed measurements of the TFT characteristics. Pulsed measurements are expected to minimize the stress effects, while quasi-static measurements allow stress effects to develop, which may distort the data. It is interesting to understand how the stress effect influences the apparent mobility.

Figure 2 shows a typical family of pulsed TFT measurements at temperatures between 90 K and 300 K. The TFT sub-threshold slope decreases and the apparent threshold voltage shifts to larger bias with decreasing temperature. Extracting the mobility as a function of temperature depends considerably on the analysis and interpretation of the data because there is increasing deviation from the linear relation between current and voltage at low temperature. The wide sub-threshold region implies that a large fraction of holes are in low mobility states, presumably because of disorder-induced localized states.[13]

Despite the difficulty in analyzing the TFT data, it is apparent from Fig 2 that in the region of 220 K the temperature dependence is less than at either higher or lower temperatures. We extracted approximate mobility values for the most mobile carriers, by a linear fit to the high gate voltage region of the data and a comparison of pulsed and quasi-static measurements is shown in Figure 12. We use this as an approximation to the mobility to compare pulsed and quasi-static data, and not as a complete analysis. Above about 250 K, the dc and pulsed measurements give essentially identical results. This agrees with the observation that the bipolaron stress effects are quickly equilibrated at room temperature, so that the stress gives only a 20-30% drop in current unless the measurement time is very long. At temperatures below 200 K the quasi-static measurement differs from the pulsed measurement by about a factor 2. The difference seen at low temperature agrees with the expectation that the dissociation of the bipolaron is thermally activated, and that the equilibrium bipolaron concentration is larger at low temperature. Moreover, measurements at lower temperature require higher gate voltages thus increasing the bias stress effect in the DC measurements. Similar shapes of the mobility vs. 1/T plot were observed in a single crystallite[14] and polycrystalline thin films[15] of sexithiophene and in polycrystalline films of dihexylquarterthiophene (DH4T)[16] but are not observed in spin cast films of P3HT.[9]

Fig. 12 shows a regime between about 180K and 270K where the apparent mobility dips, and the difference between the pulsed and dc measurement is largest, reaching about a factor 5. It is clear that the dip is related to the minimum in Fig. 11, and is a stress effect. This observation is in agreement with the results of Gomes et al.[17] and Muck et al.[16] who observe that the trapping rate in alpha sexithiophene ($\alpha$-6T) and DH4T peaks around T=220 K.

An alternative representation of the temperature dependence is to plot the current at a fixed gate voltage, as shown in Figure 13 for pulsed measurements. The mobility dip is absent, but there is a pronounced change of slope at about 240K, which indicates a change in the transport properties around this temperature. Bias stress can only be partially responsible for this change of slope which requires further study to fully understand.

## 4. Discussion

### 4.1 <u>The bipolaron formation mechanism</u>

The bipolaron formation rate has an activation energy of 0.12 eV up to 250K and a constant value at higher temperature. Capture of two holes into a bipolaron involves overcoming the Coulomb repulsion barrier, which we previously estimated to be about 0.4 eV. The low activation energy suggests that the mechanism is primarily tunneling through the barrier rather than thermal excitation. In general one might expect a combination of both mechanisms, as illustrated in Figure 15, because a small thermal excitation up the barrier can significantly reduce the tunneling distance.

We first estimate the expected rate of hole capture. The probability, *p*, for capture by the combined tunneling and thermal activation process is,

$$p = \omega \exp(-E(R)/kT) \exp(-2R/R_o); \qquad E(R) = e/(4\pi\varepsilon\varepsilon_o R), \qquad (3)$$

where $\omega$ is the attempt frequency, $E(R)$ is the Coulomb potential energy (with $kT$ expressed in units of eV) and $R_o$ is the tunneling parameter. The maximum rate occurs when the exponent is minimized, which occurs for a tunneling distance given by,

$$R_M^2 = e R_o /(8\pi\varepsilon\varepsilon_o kT), \qquad (4)$$

and the effective temperature dependent activation energy is then ,

$$E_M(T) = (2ekT /\pi\varepsilon\varepsilon_o R_o)^{1/2} \qquad (5)$$

According to this expression, choosing $R_0$=1nm leads to an activation energy $E_M$(150 K)=0.22 eV increasing to 0.31 eV at room temperature. As expected, the energy of the combined process is smaller than the barrier height, but it is significantly larger than the observed activation energy. Furthermore, the calculated energy increases with temperature because activation is increasingly favored by the thermal energy, whereas the measurements find a constant energy at low temperature and no temperature dependence at higher energy.

This simple model evidently does not fully account for the data and indicates that the mechanism of bipolaron formation is more complex. Many alternative mechanisms may be involved in the capture. For example, the transition from temperature independent to thermally activated behavior could be a change from reaction limited to diffusion limited kinetics. The measurement of the capture kinetics is based on the total hole concentration in the channel. However, it is evident from the TFT characteristics that the channel holes have a wide distribution of mobilities at low temperature, presumably because of the presence of shallow traps. The temperature dependence of the rate constant may reflect the activation energy for the diffusion of trapped carriers.

Alternatively the rate may reflect that only a fraction of holes – those with the highest mobility – participate in the trapping process. We have previously shown that at T<220 K, the mobility of PQT-12 depends strongly on the gate voltage, i.e. the charge density in the channel.[13] This behavior is evident for the data of Fig. 2: the sub-threshold region of the TFTs widens as the temperature is lowered. The gate-dependent mobility is consistent with the well-known mobility edge transport model where at low temperature a large fraction of holes is trapped in the valence band tail while current is carried by the smaller fraction of mobile holes. An approximate band-tail width of 34 meV was estimated for PQT-12.[13]

The bipolaron formation rate can be recalculated by taking into account band-tail trapping, i.e. by assuming that only the free holes participate in the trapping process. The alternative processes are described by rate constants for capture of free holes by any hole (trapped, $N_{ht}$ or free, $N_{hf}$), or just the capture of pairs of free holes,

$$kN_h^2 = k_{ft}N_{hf}N_h = k_{ff}N_{hf}^2 \qquad (6)$$

The resulting formation rate, $k_{ff}$, is approximately independent of temperature over the whole range investigated here (see Fig. 15) indicating that the apparent temperature decrease of the bipolaron formation rate at T<250K might simply be due to the progressively lower availability of free carriers able to form bipolarons at lower temperature. The formation rate corrected for trapping is k=(2.1± 0.9)x10$^{-18}$ cm$^3$/s.

4.2 kinetics of bipolaron dissociation

The equilibrium bipolaron concentration, $N_{BP,EQ}$, is given by,[1]

$$N_{BP,EQ} = k\, N_h^2/b = (N_h^2/\, N_V)\exp(U/kT) \qquad (7)$$

where the first expression is derived from the kinetic equation (eq. 2), and the second from the reaction (eq. 1). Eq. 7 relates the forward and reverse rates to the effective density of valence band states $N_V$, and the bipolaron formation energy, $U$. The reverse rate is therefore given by,

$$b = k N_V \exp(-U/kT) \qquad (8)$$

The relative magnitude of $k$ and $b$ explain the temperature dependence of the stress effect. In the high temperature regime (T>260 K), the bipolarons quickly equilibrate, and so the relative change in current due to the stress effect is given by,

$$I_o/I_{ST} = N_{h0}/N_{hEQ} = N_{h0}/[N_{h0}-2N_{BP,EQ}] = [1-2N_{BP,EQ}/N_{H0}]^{-1}. \qquad (9)$$

The equilibrium density of bipolarons increases with reduced temperature because of the exp($-U/kT$) term in eq. 7, and so the stress effect is predicted to get larger, as is observed down to about 220 K (see Figs 10 and 11). The change in behavior at lower temperature occurs because the bipolaron concentration during the typical measurement interval is determined by the kinetics rather than the equilibrium. The effective formation rate decreases starting at about 230 K (see Figure 9), and by 180 K has decreased by about an order of magnitude. The decreasing rate is evident from Figure 10.

The current decay as a function of temperature was calculated using the kinetics described by eq. 2, with the experimental values of $k$, and using eq. 7 to describe $b$, with fitting parameters $N_V$ and $U$. $N_0$ is fixed by the gate voltage. From the calculated decay at $V_G$=-20 V, the ratio of the current at time zero and time 0.1 sec is obtained and compared to the results of the experimental measurement in Fig. 11. A good fit to the data is obtained with values of $N_V$ =6x10$^{22}$ cm$^{-3}$ and $U$ =120 meV.

The discrepancy by about one order of magnitude between the fitted value of $N_V$ and the calculated spatial density of thiophene rings (~5x10$^{21}$ cm$^{-3}$) is not surprising given the simplicity of our model and the uncertainty of the parameters. It must be kept in mind that we report apparent kinetic parameters as at this point it is still unknown how trapping affects bipolaron formation. Furthermore, it is reasonable to assume that there is a distribution of bipolaron formation energies rather than only a single value (see section 4.3), which would have a significant impact on the fitted $N_V$ value. The model outlined here is meant to be semi-quantitative to show that the anomalous temperature dependence of bias stress in PQT-12 is readily explained by the formation of bipolarons.

### 4.3 Longer time stress effects

The data in Fig. 3 show that after the short time stress regime that is analyzed above, there is a longer term stress effect, with the property of having a slower recovery. There are several possible explanations for the long term stress and further measurements are needed to explore this regime further. We rule out charge injection in the dielectric since PQT-12 on thermal oxide and on PECVD dielectric have nearly identical properties as shown in Fig.7. A likely explanation is that the longer stress is also a bipolaron formation process, but at sites in the material where the bipolaron binding energy is larger (causing the slower release) and the capture cross section is smaller (causing the slower formation). This is a reasonable possibility in a disordered polymer, since there is clear evidence that the hole polaron state is sensitive to the local order.[18,19]

### 4.4. Disorder effects on the stress kinetics

As discussed in the previous papers, it seems highly likely that the formation of bipolarons is not uniform within the polymer, but occurs in specific regions. However, the electrical data gives no direct structural information. Structural studies show that the regioregular poly(thiophenes) that give high mobility TFTs have a highly textured polycrystalline structure with grain size of 2-10 nm. Polaron formation is known to be suppressed in highly ordered crystalline regions of poly(thiophenes) because the hole wavefunction is more extended. It seems reasonable to expect that bipolarons will be similarly suppressed. Hence bipolaron formation may occur predominately in the disordered regions between crystal grains. Bipolaron formation is closely coupled to the structure of the polymer, since structural relaxation provides the energy to bind the bipolaron.

It is interesting that the change of slope in the temperature dependence of the current (Fig. 13), the maximum in the bipolaron formation (Fig. 11), and the change in bipolaron formation kinetics (Fig. 9) all occur at about the same temperature. These are apparently unrelated phenomena, but obviously the similar temperature suggests a link that we do not yet understand.

## 5. Conclusion

The bias stress effect has an obvious influence on the measured value of the TFT mobility, especially when derived from quasi-static measurements. In particular we find an anomaly near 250 K where the apparent mobility is non-monotonic with temperature. As explained above, this feature results from the counteracting effects of the increasing equilibrium concentration of bipolarons, but the increasing time needed to reach equilibrium. These stress effects are reduced when measured with a pulsed gate bias and also by making measurements at low gate voltage where the effect is greatly reduced, due to the $N_h^2$ dependence.

## Acknowledgements

The authors gratefully acknowledge the PARC process line for assistance in sample preparation and the Xerox Research Center of Canada for providing the PQT-12 material, and thank M. L. Chabinyc, J. Northrup, W. S. Wong, A. C. Arias, and J.-P. Lu for helpful discussions. This work is partially supported by the Advanced Technology Program of the National Institute of Standards and Technology (contract 70NANB0H3033).

**Figures**

Figure 1: TFT transfer characteristics of PQT-12 at room temperature. Two measurements are shown, repeated after a few minutes to show the absence of hysteresis.

Figure 2: TFT transfer data at different temperatures in the linear regime of operation ($V_{DS}$=-1V), showing the wider subthreshold region at lower temperature.

Figure 3: Current decay of PQT-12 at two different gate voltages, and over an extended time range, sowing that there is a fast initial decay, then a plateau and finally a longer time decay.

Figure 4: Drain current decay as a function of time at different temperatures over an extended time range ($V_G$=-50 V), showing that the bias stress effect is temperature-dependent.

Figure 5: Transfer characteristics ($V_D$=-1V) before and after stressing the TFT with $V_G$=-50V for 7200s. Linear scale (a) and semilogarithmic scale (b).

Figure 6: Recovery of the switch-on voltage $V_{on}$ after stressing a TFT at $V_G$=-50V for t=7200s. Recovery after stressing for t<100s takes a few seconds (no threshold voltage shift is observed).

Figure 7: Current decay of PQT-12 at $V_G$=-20V, $V_{DS}$=-1V on thermal silicon oxide and PECVD silicon oxide/nitride dielectric.

Figure 8: (a) Drain current decay data as a function of gate voltage, showing that the decay rate is higher at higher gate voltage. (b) Charge removal rate measured from data as in (a), plotted as a function of $N^2$. The slope of the line is the bipolaron formation rate-constant.

Figure 9: Bipolaron formation rate as a function of inverse temperature. The break in the curve occurs at 220K

Figure 10: The decrease in TFT current after applying $V_G$=-20V, measured (a) and calculated (b) at different temperatures. Data are normalized to unity and offset vertically.

Figure 11: Comparison of current decay calculated through the bipolaron kinetic model with the experiment. The bipolaron formation constant used in the model is the one measured experimentally. $N_0$ is given by the gate voltage ($V_G$=-20V) and a channel thickness of 0.5-1 nm. The fit parameters are $N_V$ (available hole states) and U (bipolaron binding energy).

Figure 12: Mobility measured in DC and pulsed as a function of temperature. The pulsed mobility is measured at t=0 by subtracting the capacitive charging current from the actual current. Stress effect is maximum between 180K and 280K.

Figure 13: Drain current as a function of 1/T at different gate voltages. There is a break in the temperature dependence at 240K

Figure 14: Coulomb potential model for hole capture to form a bipolaron, as discussed in the text.

Figure 15: Comparison of the apparent bipolaron formation rate with the formation rate corrected for trapping of holes in band-tail states.

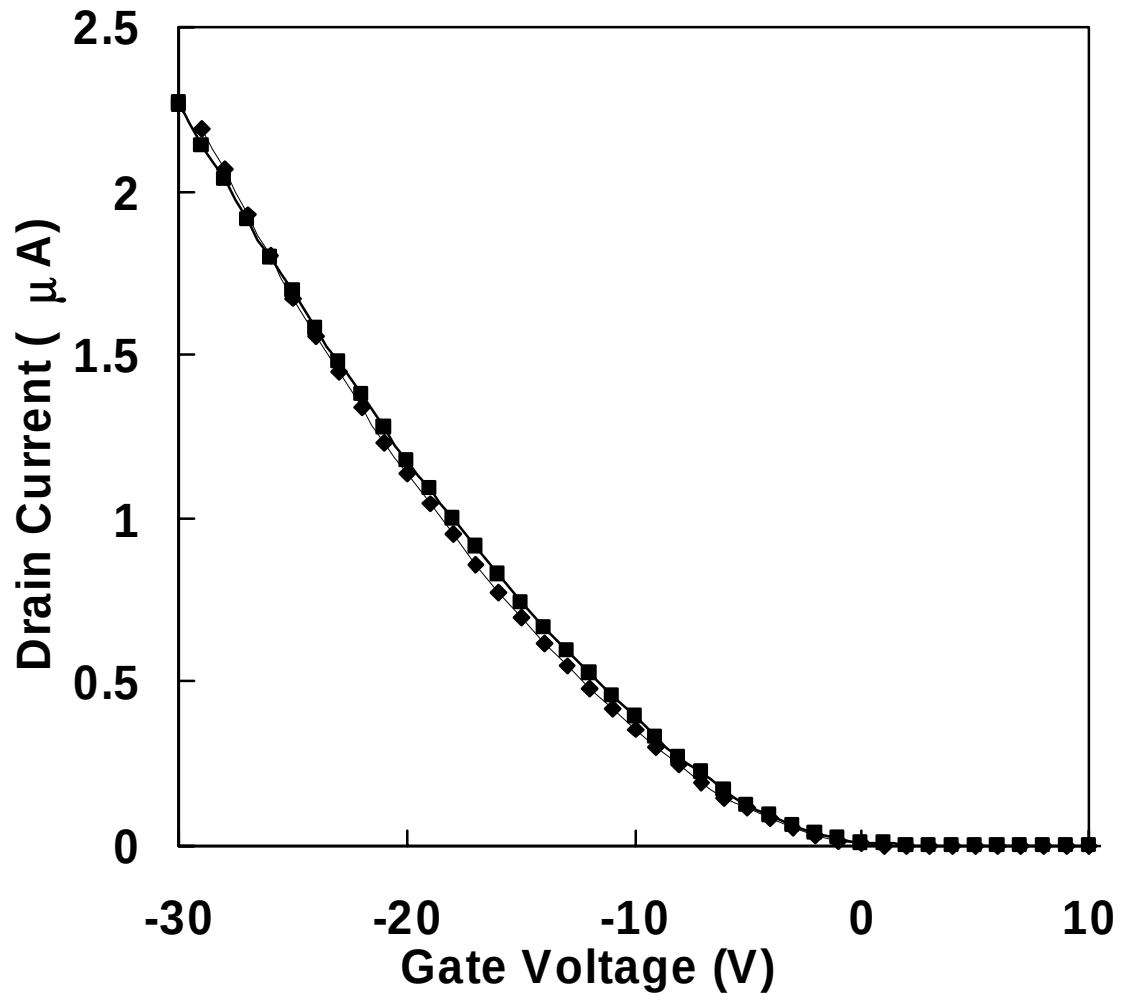

Figure 1

A. Salleo and R. A. Street: Kinetics of bias stress and bipolaron formation …

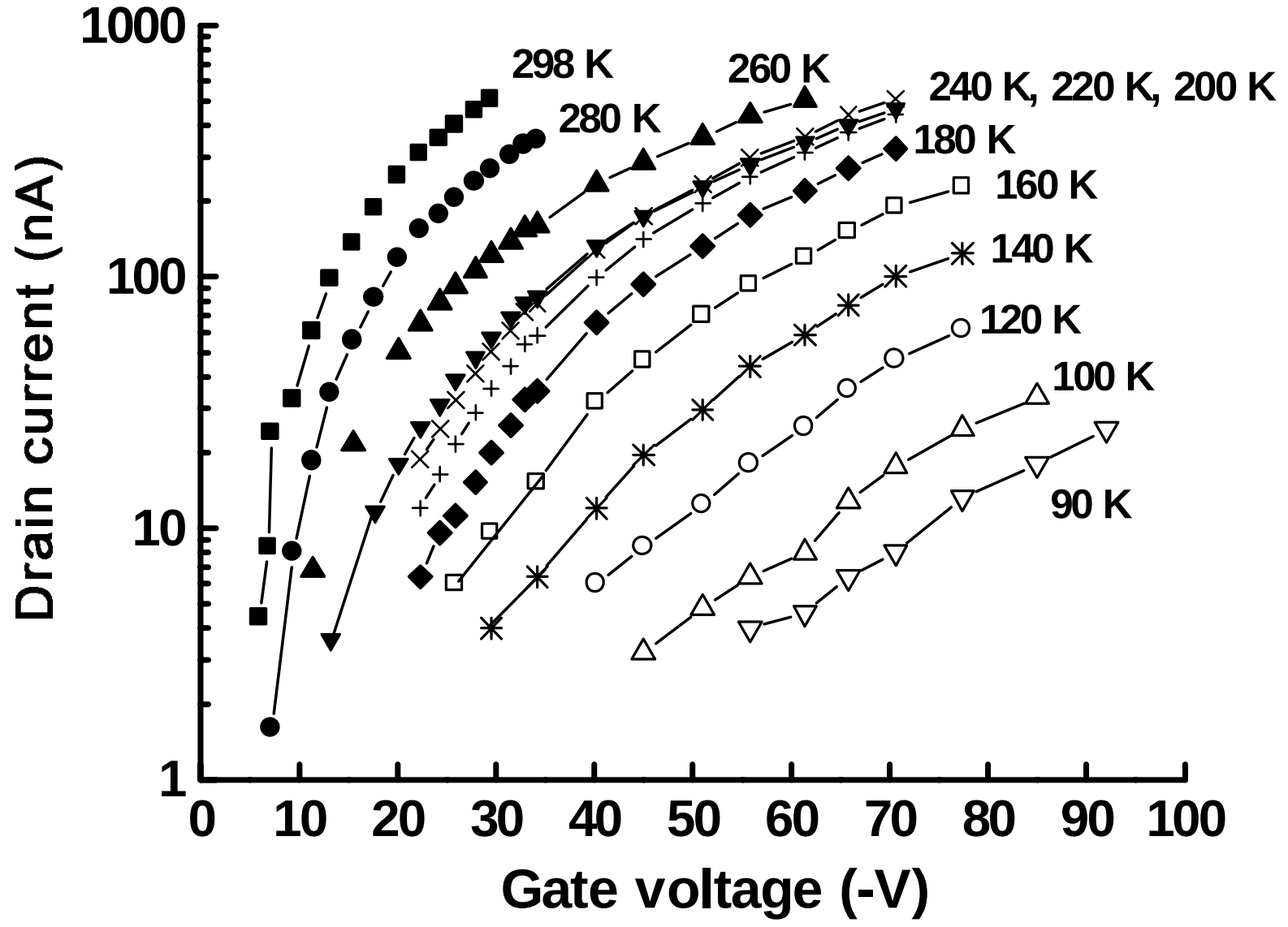

Figure 2

A. Salleo and R. A. Street: Kinetics of bias stress and bipolaron formation …

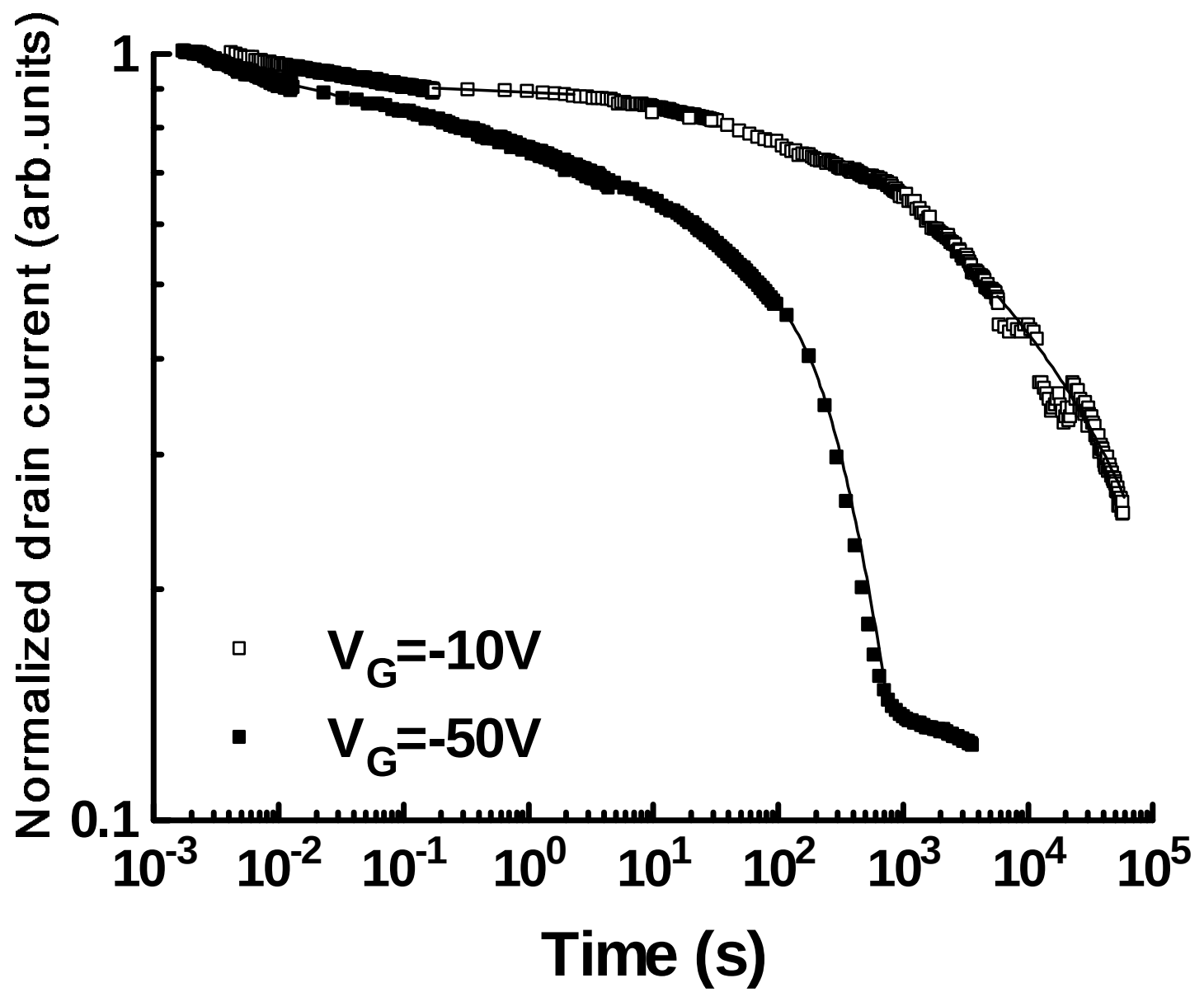

Figure 3

A. Salleo and R. A. Street: Kinetics of bias stress and bipolaron formation …

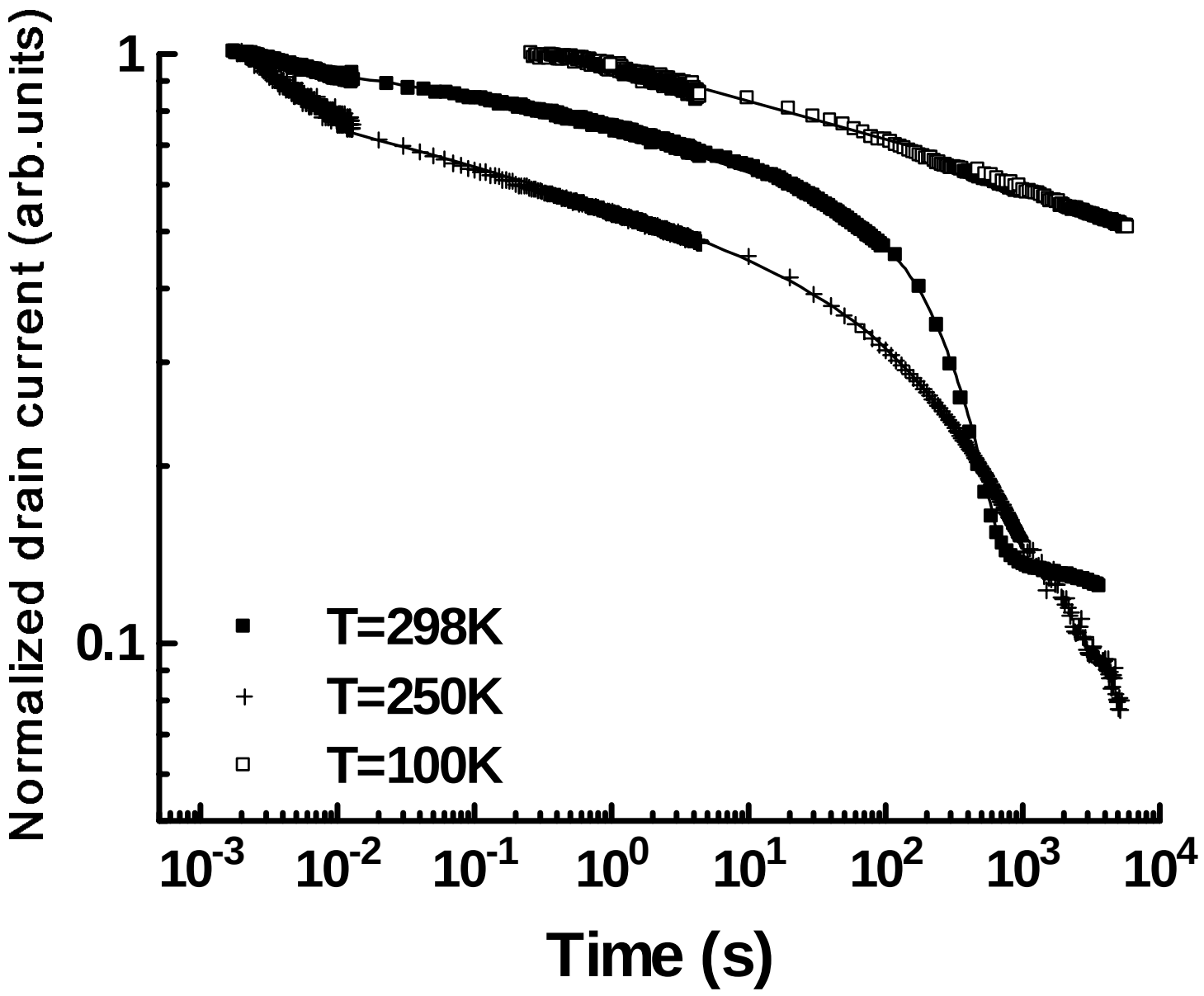

Figure 4

A. Salleo and R. A. Street: Kinetics of bias stress and bipolaron formation …

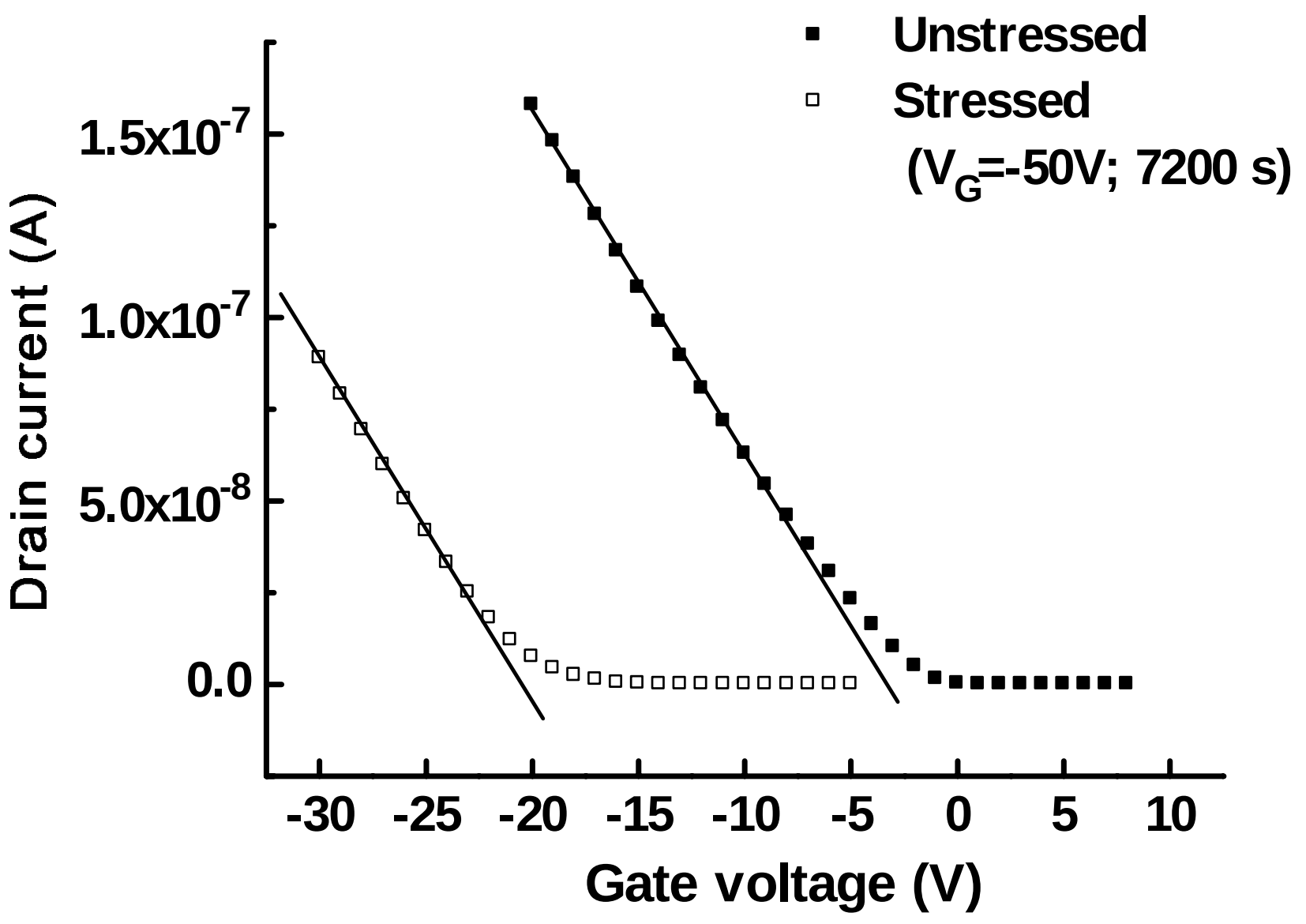

Figure 5a

A. Salleo and R. A. Street: Kinetics of bias stress and bipolaron formation …

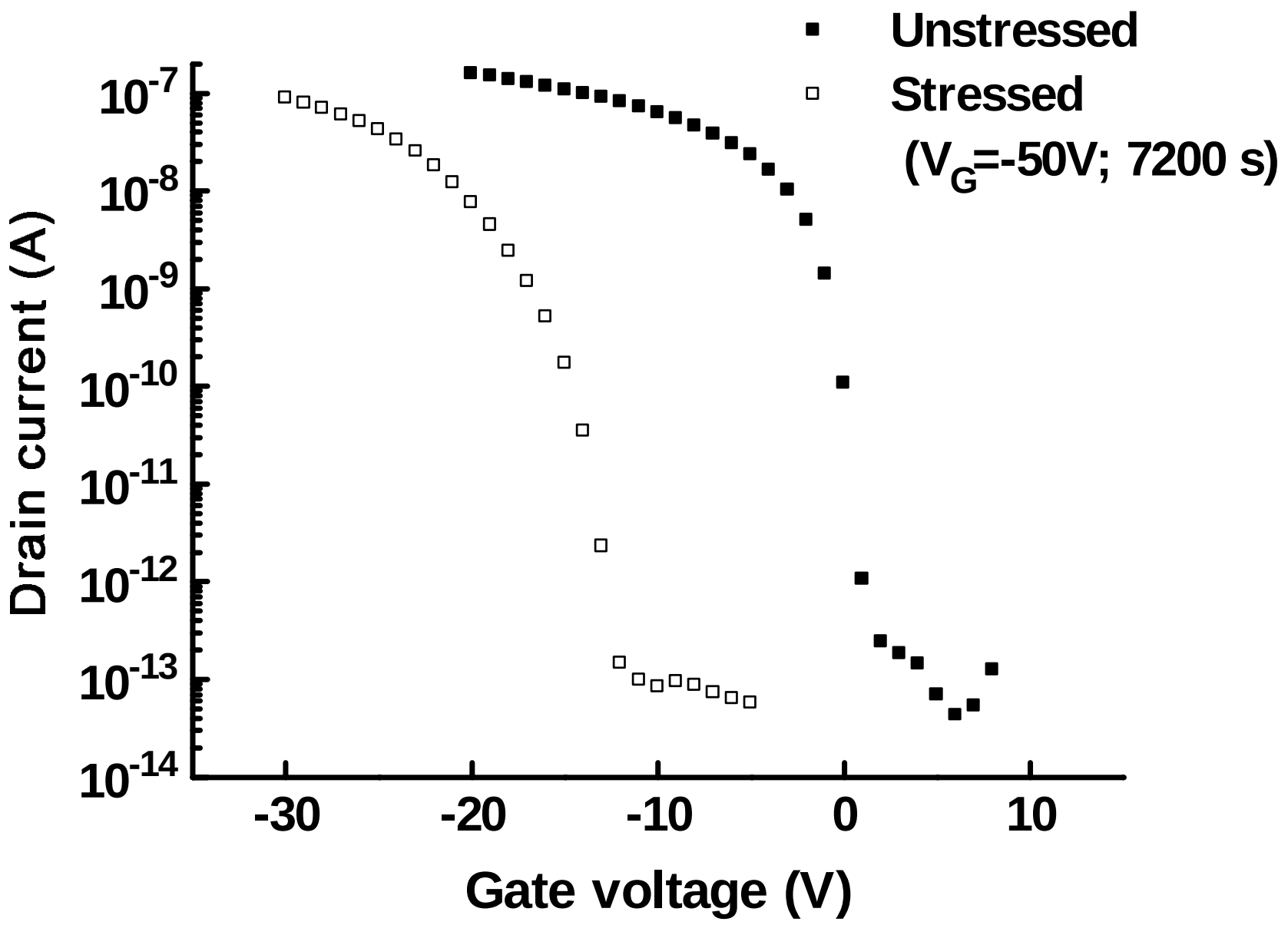

Figure 5b

A. Salleo and R. A. Street: Kinetics of bias stress and bipolaron formation …

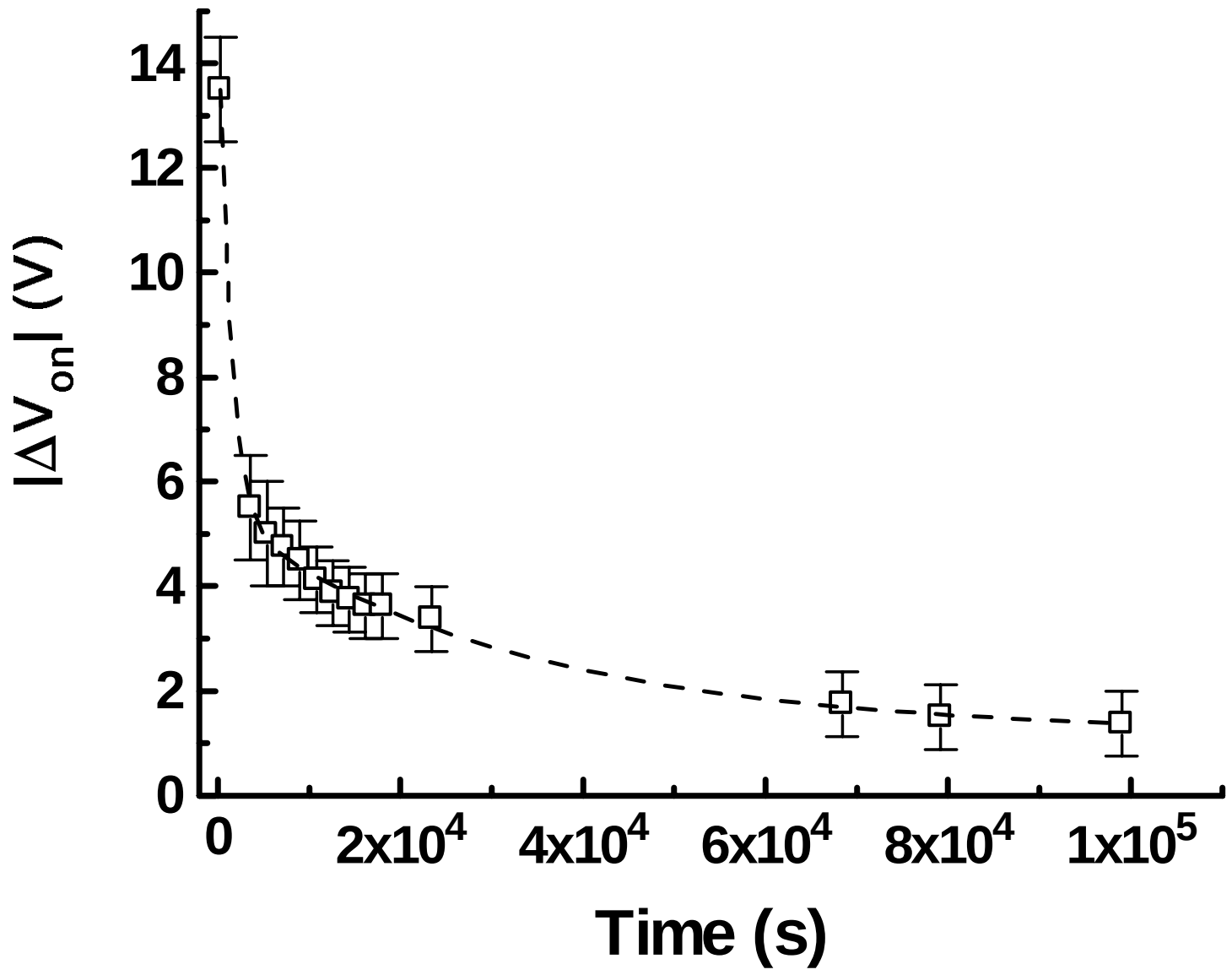

Figure 6

A. Salleo and R. A. Street: Kinetics of bias stress and bipolaron formation …

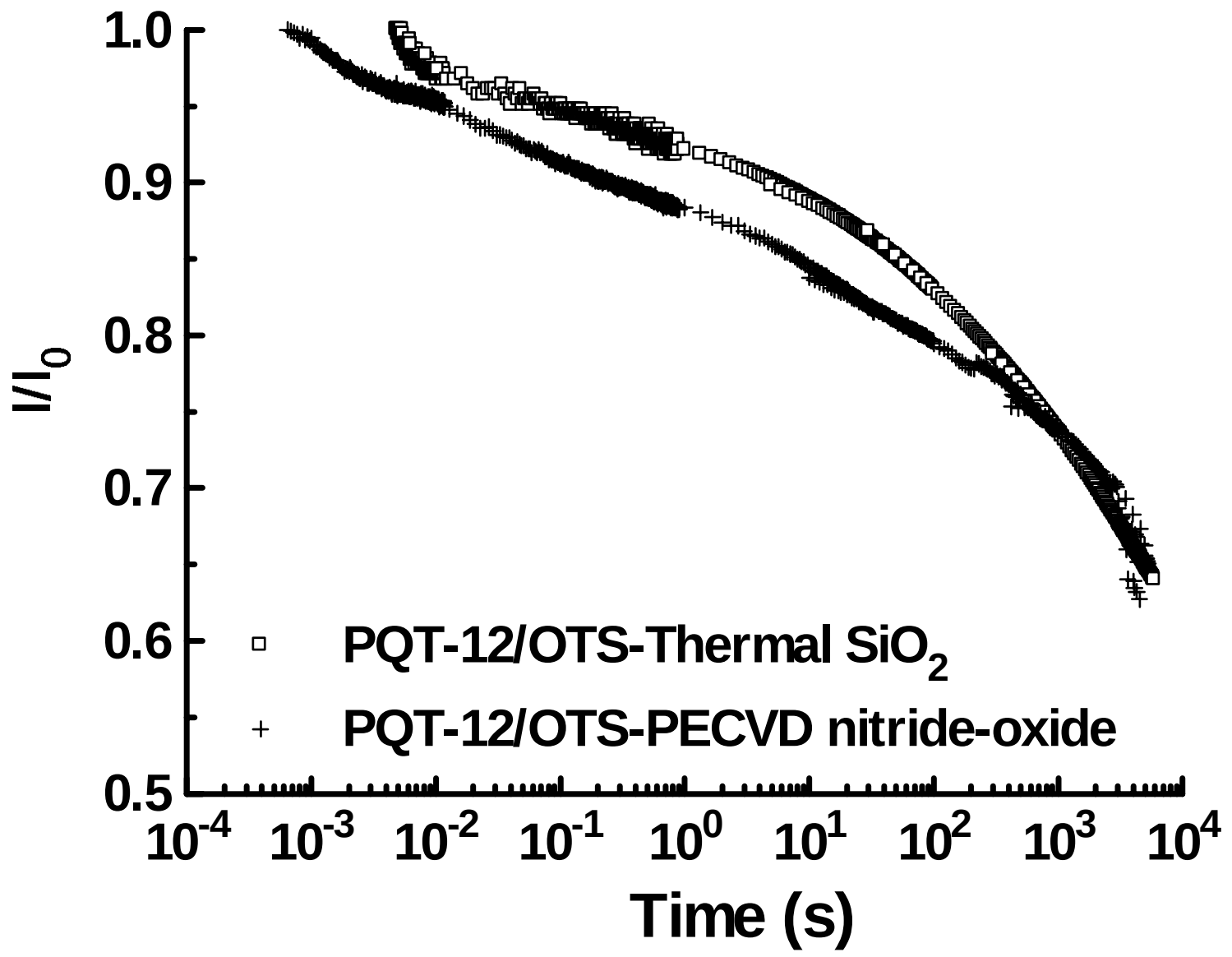

Figure 7

A. Salleo and R. A. Street: Kinetics of bias stress and bipolaron formation …

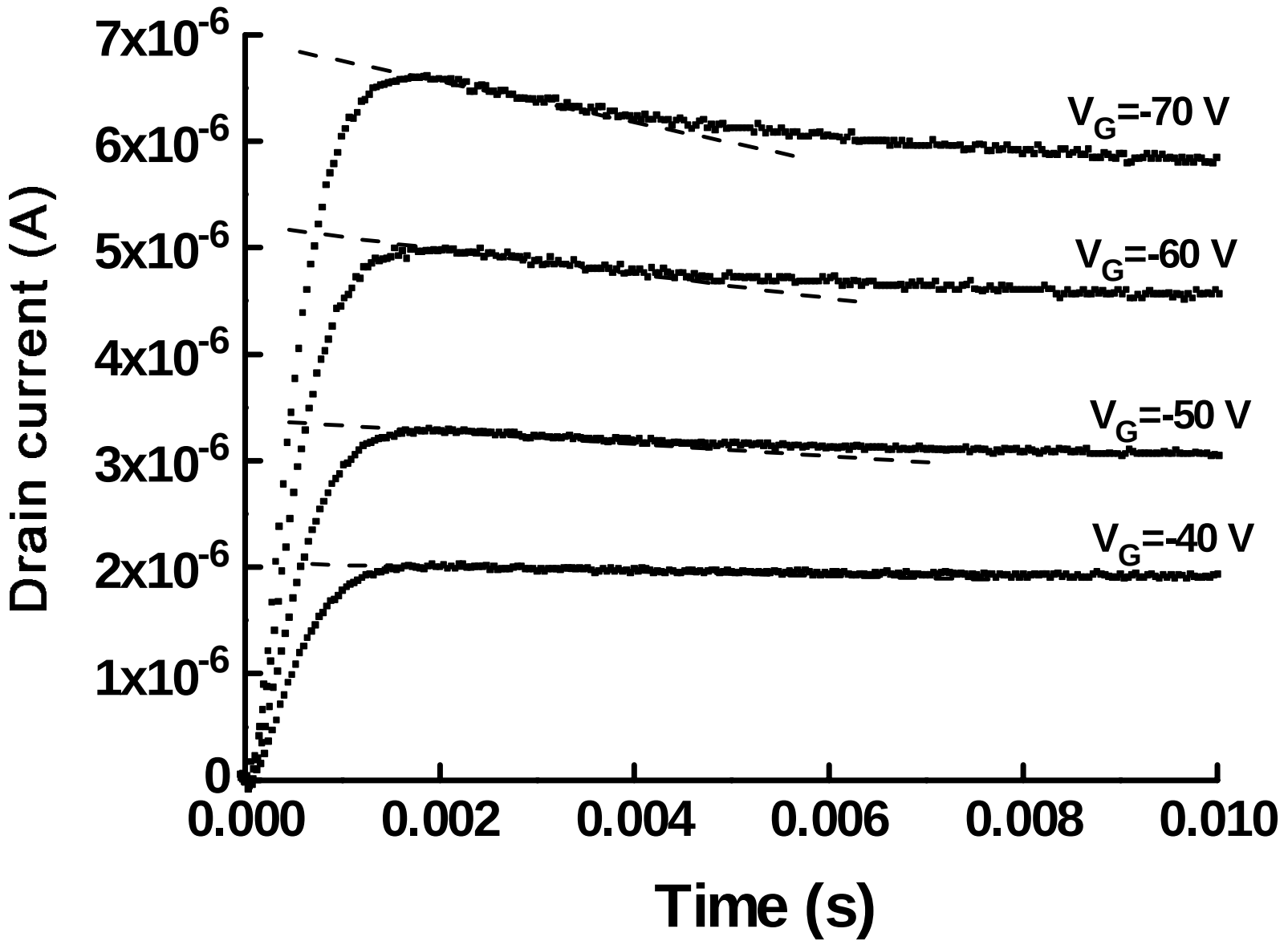

Figure 8a

A. Salleo and R. A. Street: Kinetics of bias stress and bipolaron formation …

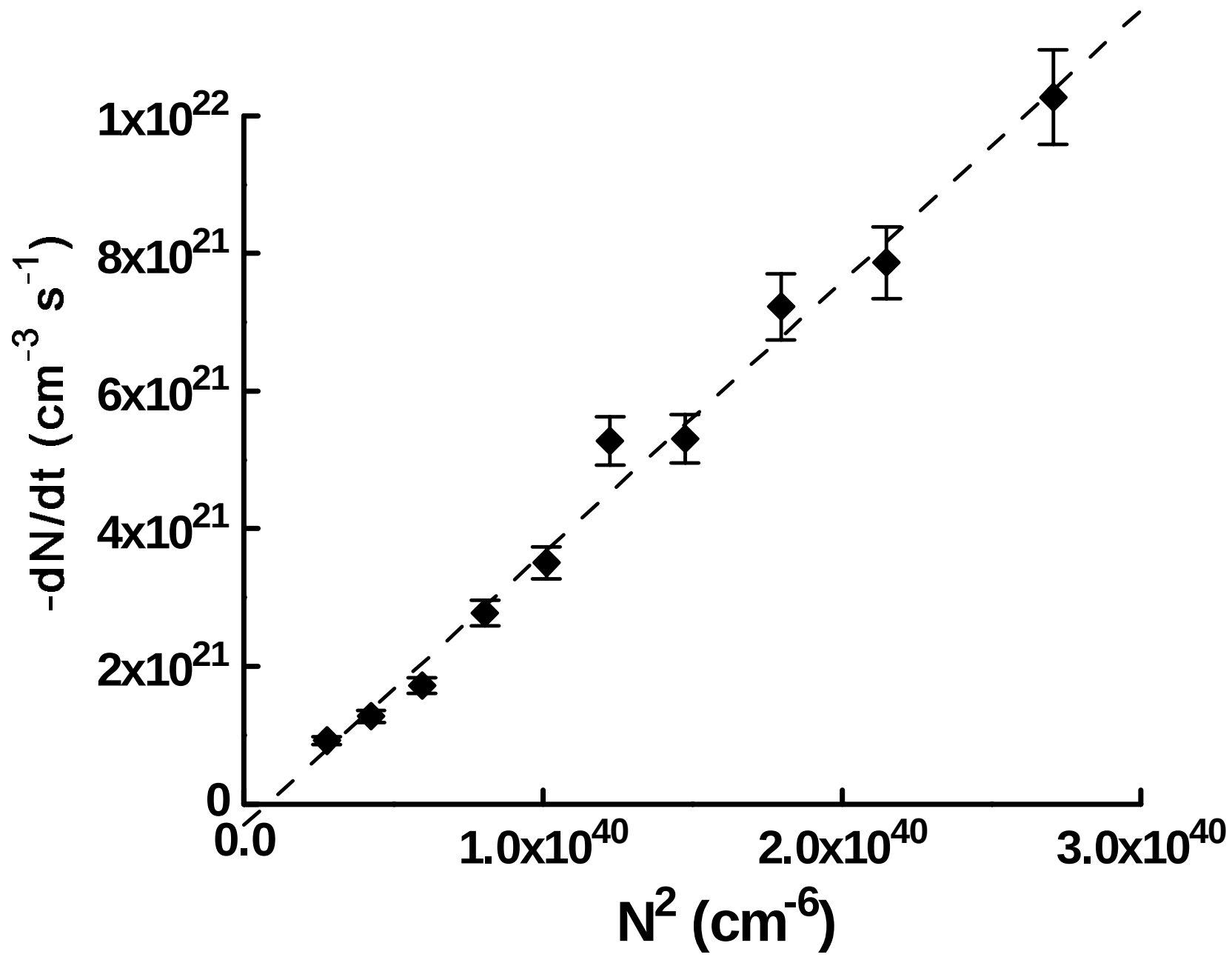

Figure 8b

A. Salleo and R. A. Street: Kinetics of bias stress and bipolaron formation …

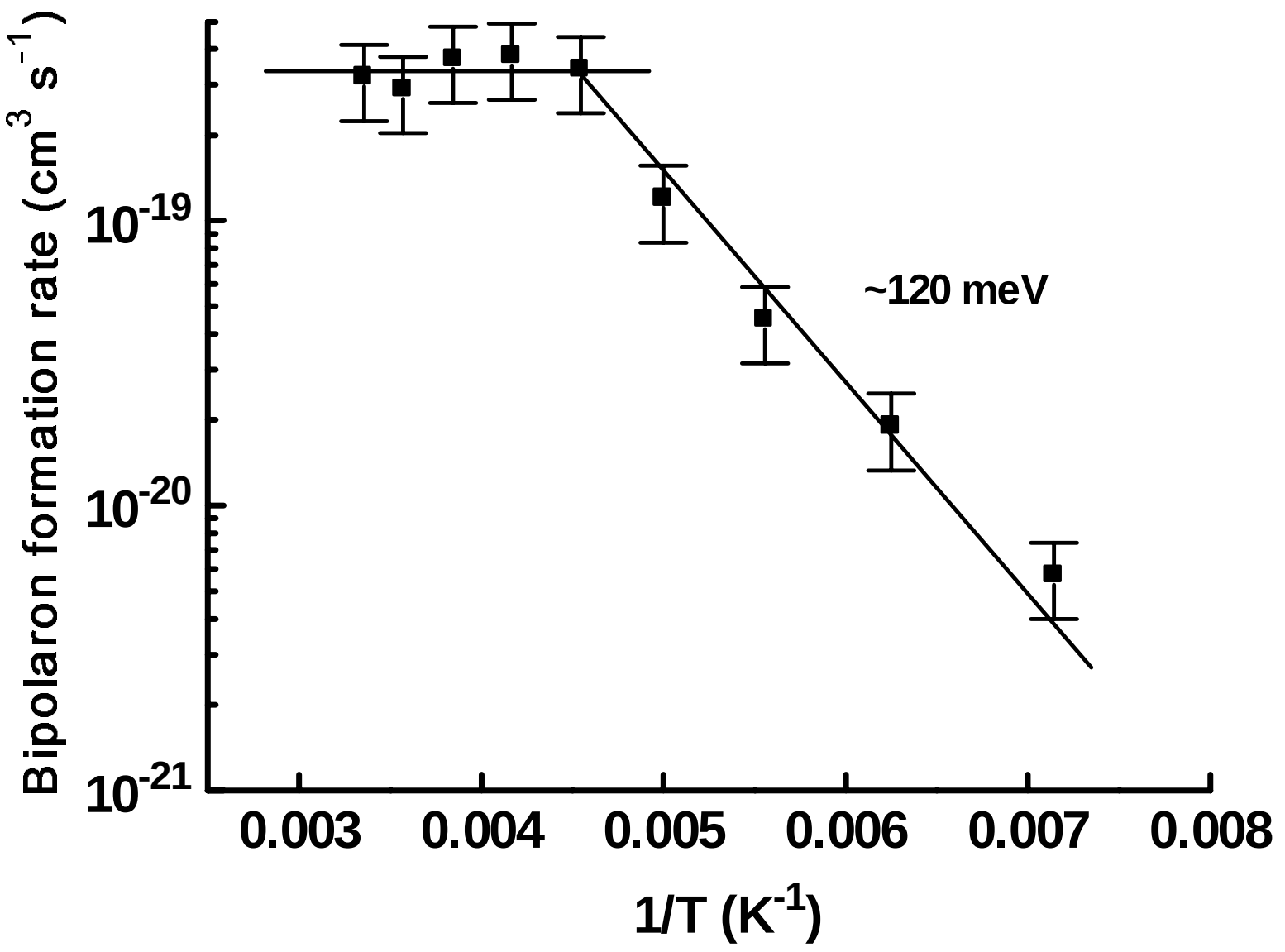

Figure 9

A. Salleo and R. A. Street: Kinetics of bias stress and bipolaron formation …

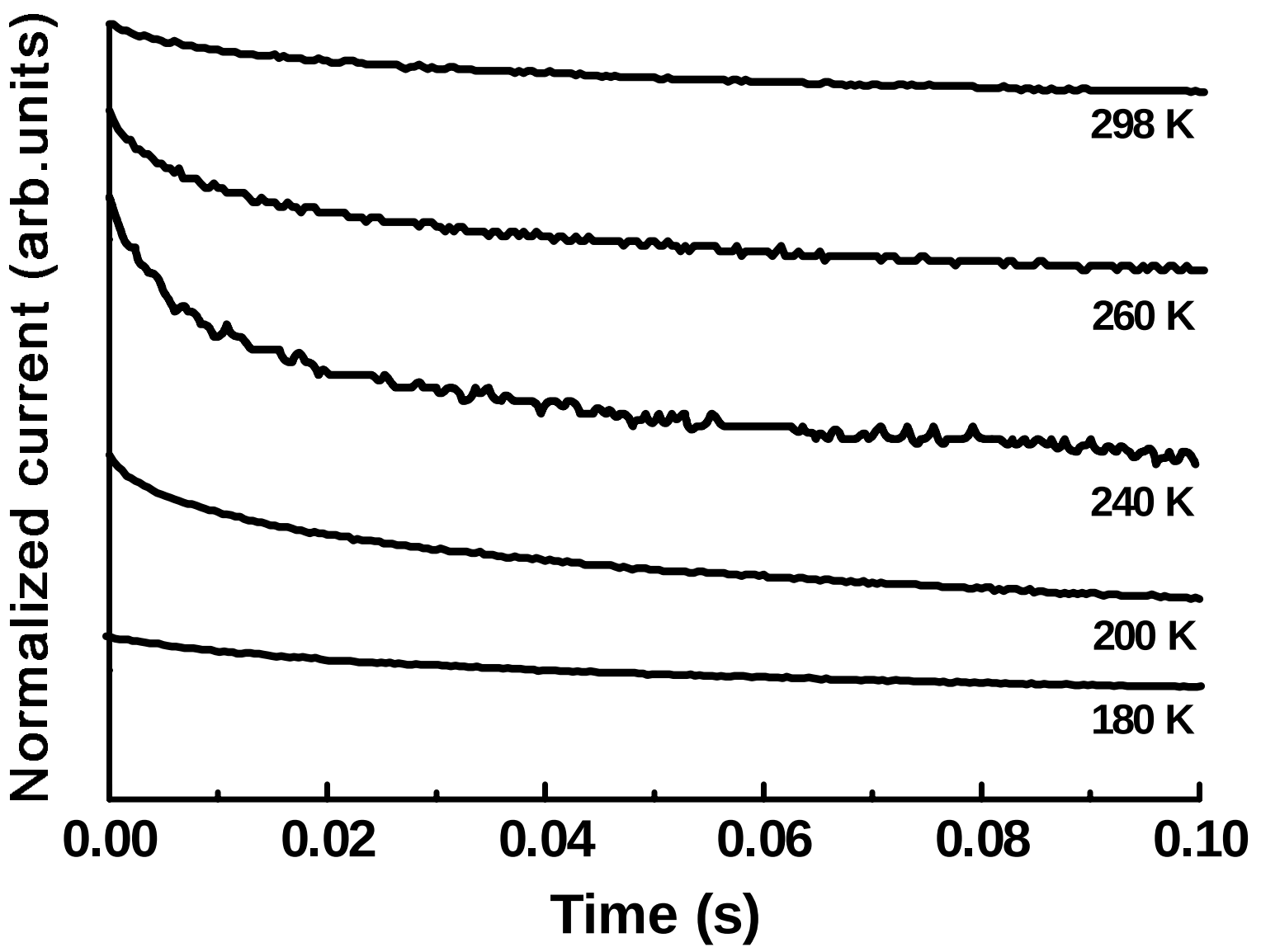

Figure 10a

A. Salleo and R. A. Street: Kinetics of bias stress and bipolaron formation …

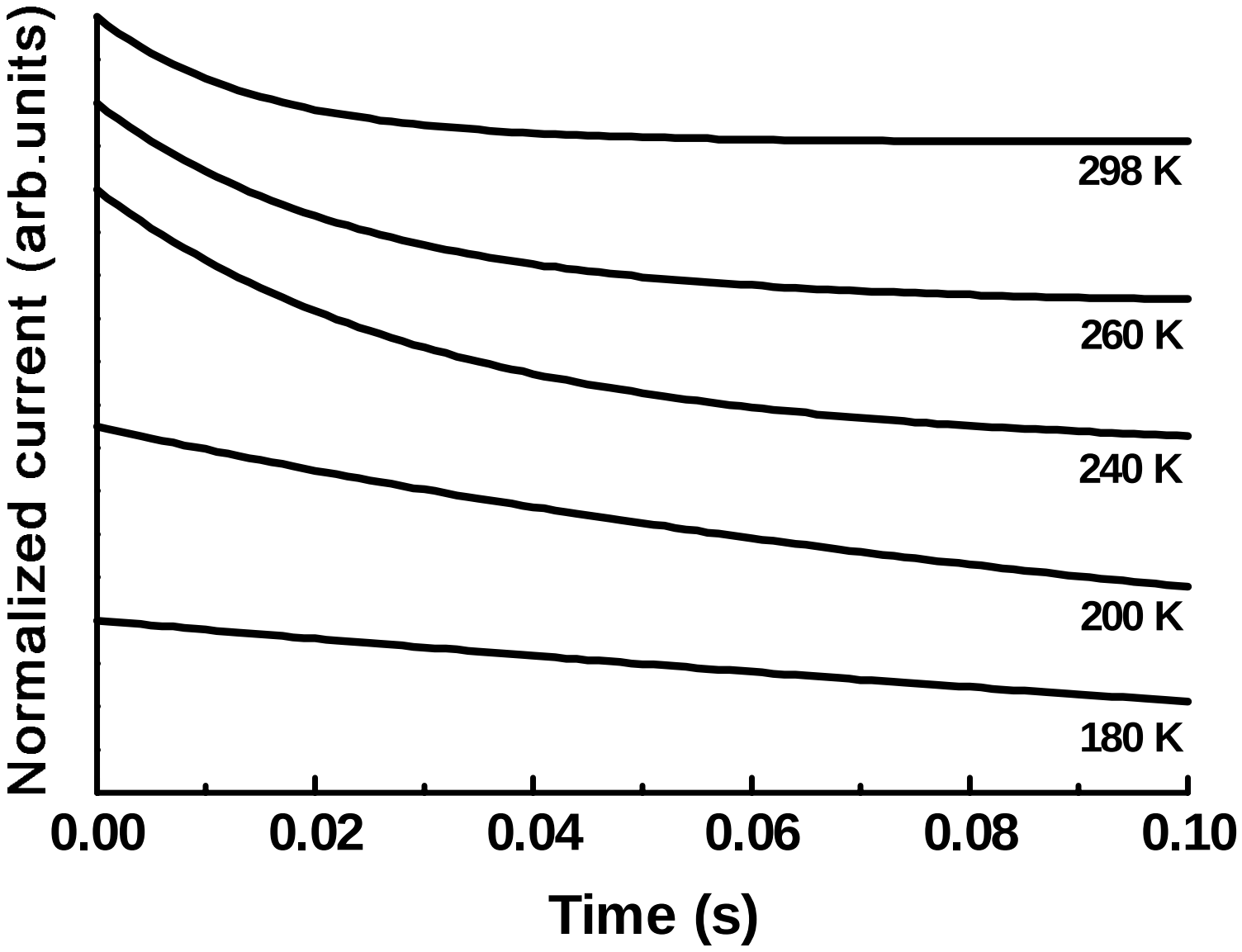

Figure 10b

A. Salleo and R. A. Street: Kinetics of bias stress and bipolaron formation …

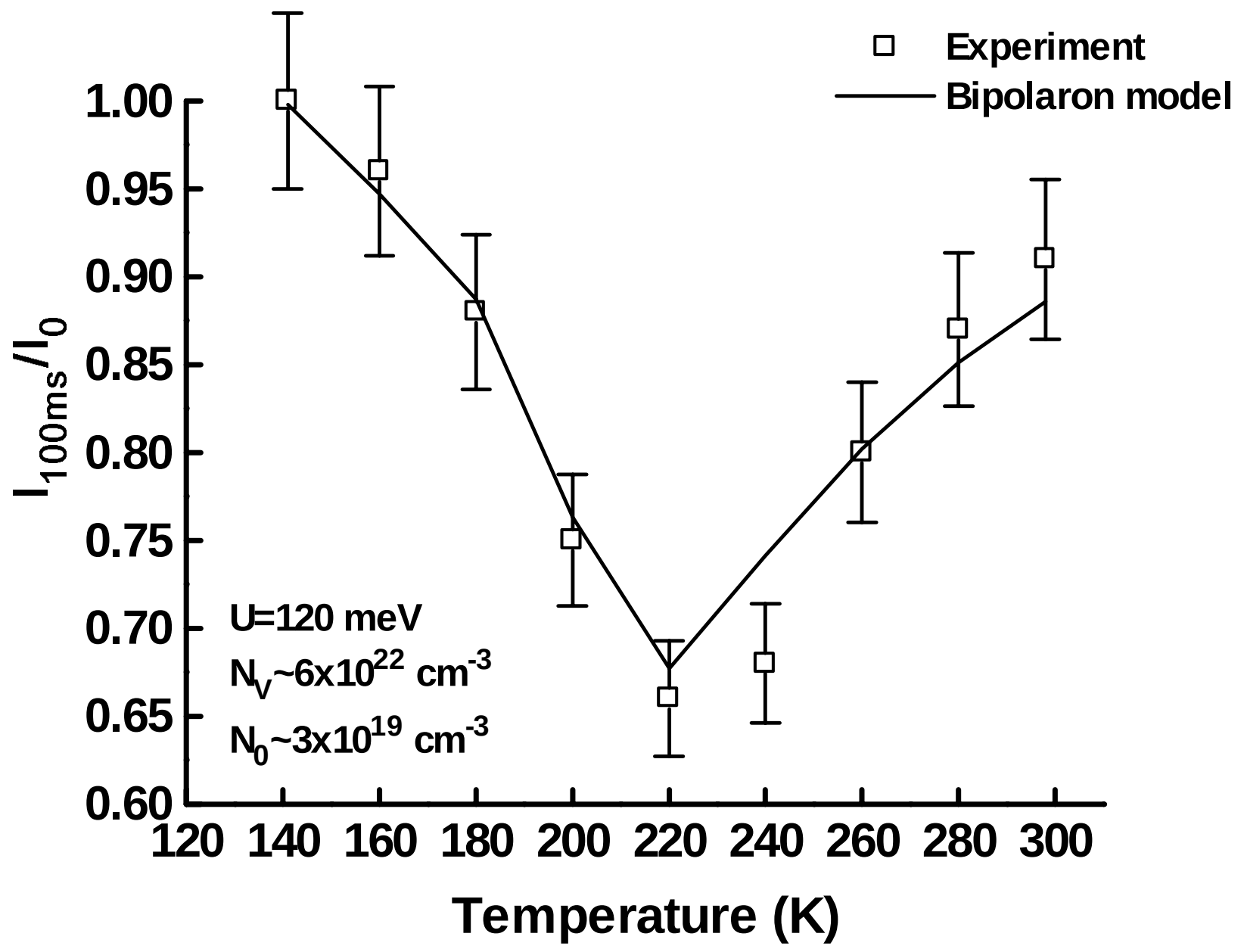

Figure 11

A. Salleo and R. A. Street: Kinetics of bias stress and bipolaron formation …

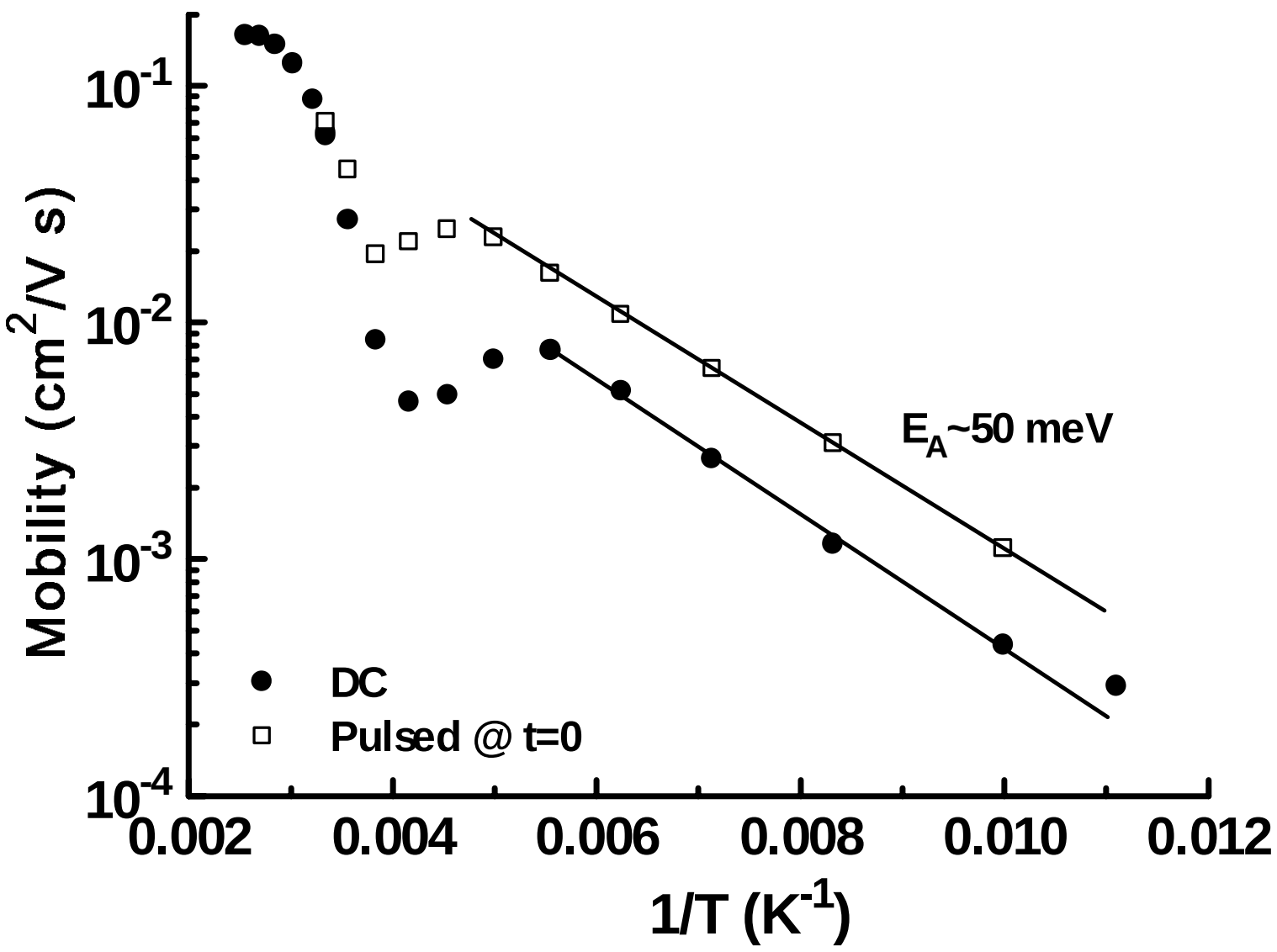

Figure 12

A. Salleo and R. A. Street: Kinetics of bias stress and bipolaron formation …

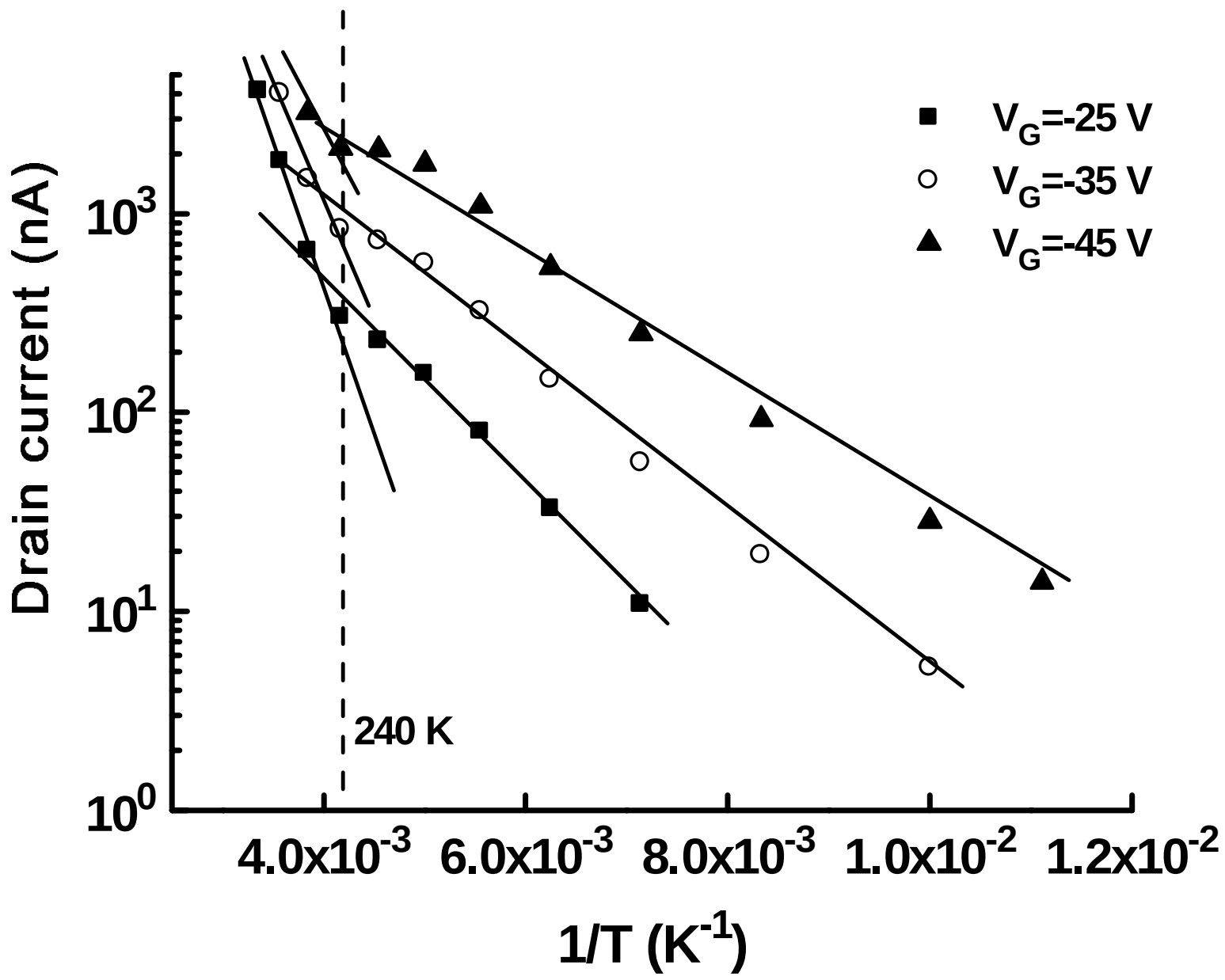

Figure 13

A. Salleo and R. A. Street: Kinetics of bias stress and bipolaron formation …

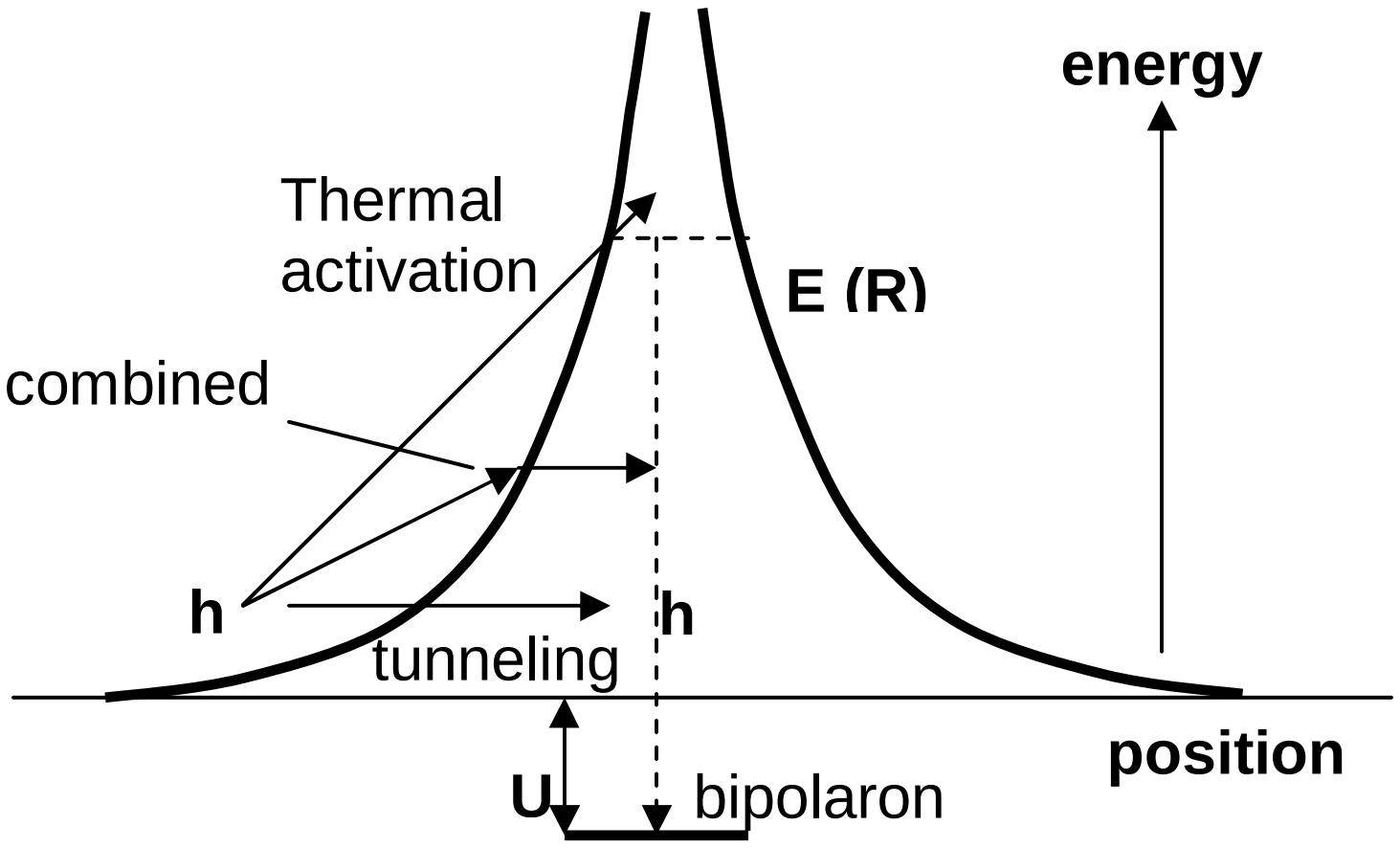

Figure 14

A. Salleo and R. A. Street: Kinetics of bias stress and bipolaron formation …

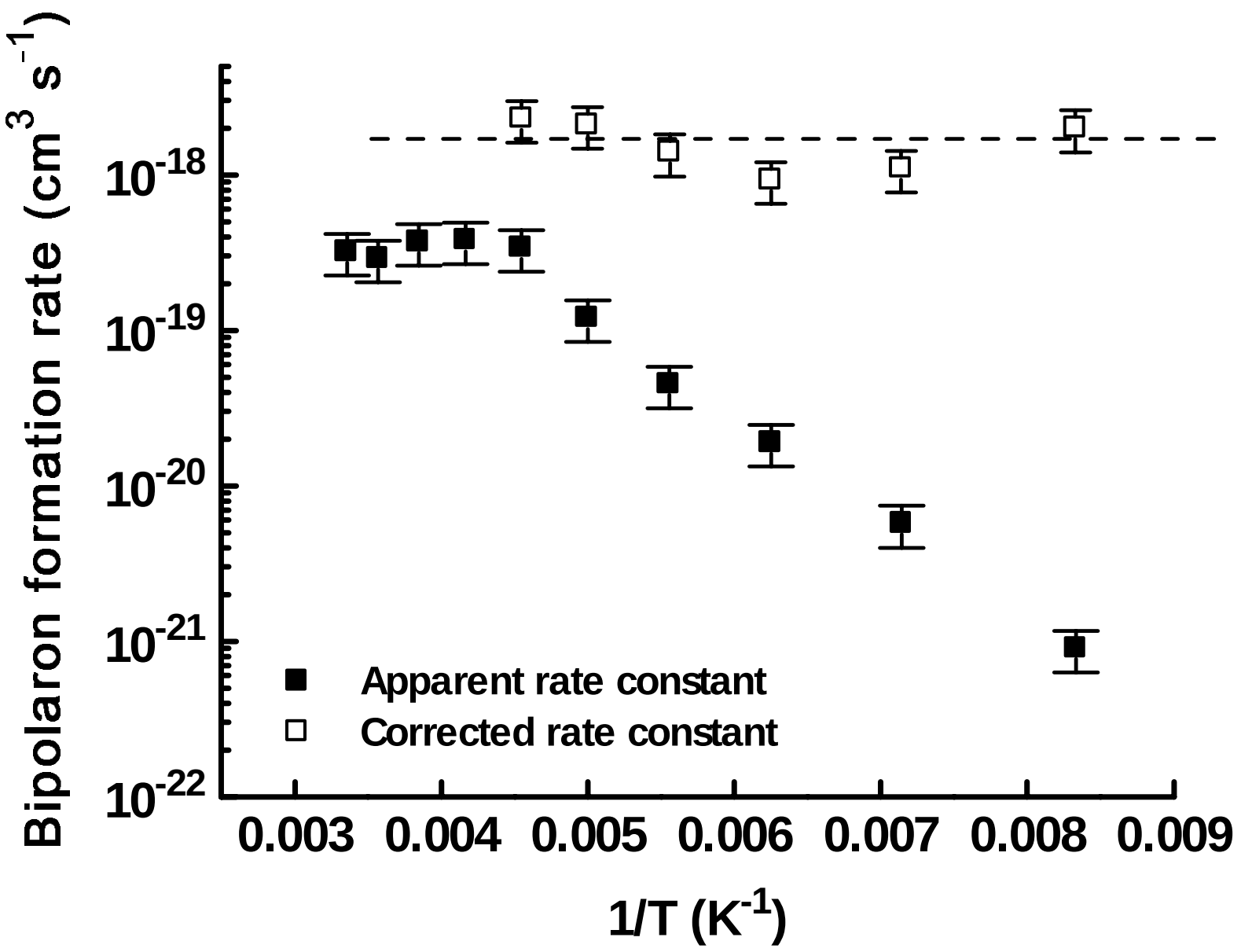

Figure 15

A. Salleo and R. A. Street: Kinetics of bias stress and bipolaron formation …